\documentstyle[aps,preprint]{revtex}
\newcommand{\be}{\begin{equation}}
\newcommand{\ee}{\end{equation}}
\newcommand{\bea}{\begin{eqnarray}}
\newcommand{\eea}{\end{eqnarray}}

\begin{document} 

\baselineskip=24pt
\begin{center}
{{\bf 
{\large\bf Time-Dependent Variational Analysis of Josephson
Oscillations in a Two-component Bose-Einstein Condensate }  
}} 
\end{center}
\begin{center}
Chi-Yong. Lin$^{1}$, E. J. V. de Passos$^{2}$
\end{center}
\begin{center}
{\it
Instituto de F\'{\i}sica, Universidade de S\~ao Paulo, CP 66318\\  
CEP 05389-970, S\~ao Paulo, \ SP, \ Brazil}
\end{center}

%\vskip 0.3cm
\begin{center}
Da-Shin Lee$^{3}$
\end{center}
\begin{center}
{\it
Department of Physics \ National Dong Hwa University \\
Hua-Lien, Taiwan, \ R.O.C.}
\end{center}

\vskip .3cm
\baselineskip=18pt
\begin{center}
{\bf ABSTRACT}
\end{center}
\vskip 0.1cm
\indent{The} dynamics of Josephson-like oscillations between two
coupled Bose-Einstein condensates is studied using the time-dependent
variational method.  
We suppose that the quantum state of the condensates is a gaussian
wave-packet which can translate and perform breathing shape
oscillations. 
Under this hypotheses we study the influence of these degrees of freedom
on the tunneling dynamics by comparing the full-model with 
one where these degrees of freedom are ``frozen'' at
its equilibrium values.
The result of our calculation shows that when the traps are not
displaced the two models agree, whereas when they are, the
models differ considerably, the former being now closer to its linear
approximation. 

%In our study, we also investigate, in the context of the full-model,
%under what conditions the linear dynamics is valid. 

\vskip 1.5cm

\vfill
\hspace{\fill}

\noindent\makebox[66mm]{\hrulefill}

\footnotesize 
\baselineskip=12pt
\indent$^{1}$e-mail: lcyong@fma.if.usp.br \\
\indent$^{2}$e-mail: passos@fma.if.usp.br \\
\indent$^{3}$e-mail: dslee@cc3.ndhu.edu.tw \\
\newpage
\normalsize
\baselineskip=24pt

\bigskip

\indent{The}
investigation of collective excitations of Josephson-coupled
two-component Bose-Einstein condensates has been the subject of many
papers \cite{Sm97,RSK99,Wi99}.
The main interest in those papers is the study of the dynamics
associated to the exchange of atoms between the two condensates.
The basic hypotheses in most of these studies is that this exchange is
``coherent''\cite{Timmermans98}, that is, without changing the quantum state of each
condensate, which leads to a dynamics involving only the degree of
freedom associated to the relative population and phase of the
condensates. 
Based on this ``frozen'' model, a host of interesting
phenomena is predicted, for example, the existence of
a ``phase-transition'' to a non-symmetrical equilibrium composition in
the limit of strong repulsion between the two condensates and ``weak''
Josephson coupling\cite{Sm97,RSK99}. 

\indent{One}
exception, however, is the work of reference \cite{Wi99} where the
effect of the change of the quantum state is taken into account. 
They use standard mean-field theory to derive coupled time-dependent
Gross-Pitaevskii equations for the time-evolution of the order
parameter of each condensate.
They simplify the problem by treating the system in only one spatial
dimension and they observe the effect of the mean-field on the
non-linear oscillations of the relative population between the
two condensates. 

\indent{In}
this paper we study the effect of the change of the quantum state in
the tunneling dynamics. 
Our approach is in the spirit of reference \cite{Wi99}, however we take
a more qualitative point of view along the  line of references
\cite{KL95,BP96,PG96,Bu97}.  
We suppose that the quantum state of the system is a gaussian
wave-packet which can translate and perform breathing shape
oscillations. 
Under this hypotheses, we study the influence of these degrees of freedom
on the tunneling dynamics by comparing the full-model with 
the limit where these degrees of freedom are ``frozen'' at
its equilibrium values.
In our study, we also investigate, in the context of the full-model,
under what conditions the linear dynamics is valid. 

\indent{Our}
starting point is the action 
\be
S \;=\; \int d^3 {\vec r} dt\;
\Bigl[\;\; \sum_{j} i\hbar \psi^{*}_{j} (\vec r,t) \dot{\psi}_{j}
(\vec r,t) - {\cal E} (\vec r,t)\; \Bigr]
\ee
 
\noindent{where} 
$\psi_{j} (\vec r, t)$, $j=1,2$ are the order parameters of each
condensate and the energy density is given by
\be
%{\cal H}&=&{\cal H}_{c} + {\cal H}_{jos} \nonumber\\
{\cal E}(\vec r,t)=
\sum_{j}
\psi_j(\vec r,t)^*\Bigl[-\frac{\hbar^2\nabla^2}{2m}
     +V^{j}_{trap}+\frac{\delta}{2}\gamma_j\Bigr]\psi_j 
+\frac{1}{2}\sum_{k,j}\lambda_{kj}|\psi_j(\vec r,t)|^2|\psi_k(\vec r,t)|^2 
+\Omega \Bigl[\psi^*_1\psi_2+\psi^*_2\psi_1\Bigr] \;.
\ee

\noindent{In}
the above expression $V^j_{trap} (\vec r)$ is the trapping potential of
each component which we take as
\be
V^j_{trap}=\frac{m}{2}\Bigl[\omega^2_{\perp}(x^2+y^2)+\omega^2_z(z+\gamma_j
z_0)^2 \Bigr]
\ee  
 
\noindent{where}
$\gamma_1=1$ and $\gamma_2=-1$,
$\lambda_{jk}$ are  proportional to the scattering
lengths $a_{jk}$ for the collisions of atoms $j$ and $k$ of equal mass, 
$\lambda_{jk}=\frac{4\pi}{m}a_{jk}$, the term that depends on $\delta$
takes into account the effects of detuning, and the last term is the
Josephson interaction term where $\Omega$ is the tunneling intensity
\cite{Wi99}.  

\indent{The}
condition that the action is stationary with respect to variations of
the order parameters leads to the coupled mean-field equations for
$\psi_j(\vec r, t)$ \cite{Wi99}. 
In this paper we follow references \cite{KL95,BP96,PG96,Bu97} and instead we
parametrize the time dependence of the order parameter through the
variation of a set of $2n$ parameters \cite{KK76} which we denote by
${\bf X} = \{ X_1, X_2,..., X_{2n} \}$, 
\be
\psi_j(\vec r, t) = \psi_j(\vec r, {\bf X} (t))
\ee

\noindent{Given}
this parametrization the action reduces to a ``classical action''in terms 
$X_i$(t) \cite{KK76}
\be
S \;=\;\int dt \; \Bigl[ \sum_{k}\Gamma^{(1)}_{k}({\bf X}) 
\dot X_{k} - E({\bf X}) \Bigr]
\ee

\noindent where 
\be
\Gamma^{(1)}_k\;=\;\frac{i}{2}\sum_{j}\left( 
\langle\psi_j | \frac{\partial \psi_j}{\partial X_k}\rangle
-\langle\frac{\partial \psi_j}{\partial X_k} | \psi_j\rangle
\right) 
\ee

\noindent{In}
(6) we introduced the state vector $|\psi_j(X(t))\rangle$ whose norm
is equal to the population of the component $j$ of the condensate and
whose wave function is given by equation (4), and $E({\bf X})$ is the spatial
integral of the energy density 
\be
E({\bf X})=\int d^3 \vec{r} {\cal E}(\vec r,t)
\ee

\noindent{with}
the order parameter parametrized as in eq. (4).

\indent{Requiring} 
that the action is stationary with respect to variations
of the parameters leads to Hamiltonians-type equations of motion for
$X_{i}(t)$: 
\be
\sum_{k}\Gamma^{(2)}_{jk}\; \dot X_k = \; \frac{\partial E}{\partial X_j}\;,
\ee 
 
\noindent{where}
the antisymmetric matrix $\Gamma^{(2)}_{jk}$ is given by 
\be
\Gamma^{(2)}_{jk}=\frac{\partial\Gamma^{(1)}_k}{\partial X_j}
-\frac{\partial\Gamma^{(1)}_j}{\partial X_k}
\ee

\indent{To}
proceed we should specify our choice of parameters. Previous work
\cite{BP96,PG96,Bu97} have 
shown that a parametrization which reproduces very well the
equilibrium properties and collective excitations of a single-component
condensate is one where the condensate density is a gaussian
wave-packet which can translate and perform quadrupole and monopole
shape oscillations and its corresponding motions. 
The study of the two-component condensate dynamics with the above
parametrization is feasible and is under investigation.
Here, in the spirit of reference \cite{Wi99}, we simplify the problem by
treating the system in only one spatial dimension.

\indent{Thus}
we write the order parameter as:
\be
|\psi_{j}\rangle=\sqrt{N_{j}(t)}e^{-i\theta_j}
|\phi_{j}\rangle 
\ee

\noindent{where}
we put explicitly the degree of freedom associated to the population
and phase of each condensate. 
The condensate orbitals $|\phi_{i}\rangle$ \cite{BR86} are taken as
gaussian wave-packets which can
translate and perform breathing shape oscillations,
\be
\phi_{j}(z,X)\rangle
=\Bigl( \frac{1}{\sqrt{\pi} q_j} \Bigr)^{\frac{1}{2}} \;
e^{ iP_{j}(t)\bigl[z-Q_{j}(t)\bigr] -\bigl[\frac{1}{2q^2_j(t)}
-i\frac{p_j(t)}{2q_j(t)}\bigr] \bigl[z-Q_{j}(t)\bigr]^2}
\ee

\noindent{where}
the parameters $(Q_j,P_j)$ are related to the translational degrees of
freedom and $(q_j,p_j)$ to the ``breathing'' oscillation.
Notice that in this parametrization the quantum state of each
condensate (called here the condensate orbital) $|\phi_{j}\rangle$ does
not depend explicitly on the phase and population of the condensates.

\indent{Thus}, 
we have twelve parameters which leads to a twelve dimensional
``phase space''. Since $E({\bf X})$ depends only on the difference of
phase between the two condensates, the equations of motion (8) give
that the total number of particles is conserved. 
This reduces the number of degrees of freedom to ten, which we take as
besides the $q_i, Q_i, p_i, P_i, i=1,2$, the relative phase and
relative population fraction. 
Thus, in the full-dynamics, we solve ten coupled first order
equations, given an initial configuration.

\indent{Since} 
${\rm det}\Gamma^{(2)}({\bf X})\neq 0$, the equilibrium configuration of
the system is determined by the condition
\be
\frac{\partial E}{\partial X_j}(X_0)=0.
\ee

\noindent{Linearizing}
the equation (8) in the neighborhood of the equilibrium configuration,
one has
\be
\sum_{k}\Gamma^{(2)}_{jk}({\bf 0})\dot x_k(t) = 
\sum_{k}\frac{\partial^2 E }{\partial X_j\partial X_k}({\bf 0})x_{k}
\ee

\noindent{where} 
$x$ is the displacement from equilibrium 
\be
x_k=X_k-X_{k0}
\ee

\noindent{and}
${\bf 0}$ in these equations denote that the quantities are evaluated
at the equilibrium configuration.
The solution of the coupled linear equations (13) define the small
oscillation dynamics around this point.

\indent{Finally},
the frozen model is obtained by constraining all the degrees of
freedom, except the relative population fraction and the relative phase
at its equilibrium values.
When this is done, the equations of motion (7) reduces to \cite{Sm97,RSK99}
\bea
\dot\theta
&=&\Delta+\frac{\Lambda\eta}{2}-\omega_{R} \;
\frac{\eta}{(1-\eta^2)^{1/2}}\cos\theta\\
\dot\eta&=&\omega_{R}\; (1-\eta^2)^{1/2}\sin\theta
\eea

\noindent{where}
$\theta=\theta_2-\theta_1$ and $\eta=\eta_2-\eta_1$, and in equations
(15) the time is measured in unit of $\omega_0$, with $\omega_0$ the
trap frequency. 
The parameters defining the frozen model are : 
\bea
\Delta&=&\varepsilon^{2}-\varepsilon^{1}-\delta\\
\Lambda&=&\bar\lambda_{11}+\bar\lambda_{22}-2\bar\lambda_{12}\\
\omega_{R}&=&\frac{2\Omega}{\hbar \omega_0}\int d^3\vec r \phi^*_1(\vec r, {\bf
0})\phi_2(\vec r, {\bf 0})\\
\eea

\noindent{where}
$\varepsilon^j$ is essentially the energy per-particle of each
condensate (in units of $\omega_0$)
\be
\varepsilon^{j}=
\Bigl[ \langle\phi_j|t+V^j_{trap}|\phi_j\rangle
+\frac{N}{2}\lambda_{jj}
\int d^3 \vec r |\phi_j (\vec r,{\bf 0})|^4\Bigr]/\hbar\omega_0
\ee

\noindent{and}
the $\bar\lambda_{jk}$ are
\be
\bar\lambda_{jk}=N\frac{\lambda_{jk}}{\hbar\omega_0}
\int d^3\vec r |\phi_j(\vec r,{\bf 0})|^2|\phi_k(\vec r,{\bf 0})|^2
\ee

\indent{We}
apply the formalism presented above to the system considered in
reference \cite{Wi99}, trapped $^{87}$Rb atoms in hyperfine states
$|f=1,m_f=-1\rangle$ and $|f=2,m_f=1\rangle$.
Following this reference, the initial configuration  is such that the
only degree of freedom which is not at equilibrium is the phase
difference and the detuning is chosen in such a way that the system is
initially driven resonantly \cite{Wi99}, which implies that 
$\Lambda = 0$. 

\indent{As} a first example we will discuss the case where the traps are not
displaced ($z_0=0$) and $a_{11}=a_{22}=a_{12}=a_{\rm Rb}$.
In this case we have a complete agreement between the ``frozen'' and
``full model'', indicating that the tunneling dynamics is such that
the quantum state of the condensates do not change.
This result is a consequence of the fact that, in this symmetrical
case, $q_i = q_{i0}$, $Q_i = Q_{i0}$, $p_i = P_i = 0$, $i=1,2$ is an
invariant subspace (also called maximally decoupled subspace) of the
system, that is, if the system is in this surface, it remains there at
all time \cite{PS86}.

\indent{We}
can also examine the importance of non-linear effects in this
symmetrical case. 
The linearization of the frozen model gives a 
frequency of oscillation equal to the Rabi frequency $\omega_R$, that
in our case is equal to 
$\omega_R = 0.100$, which coincides with the frequency
of the lowest energy normal mode of equations (13),
$\omega^{(1)}_{nm}=0.100$.  
When compared to the full model, the linear approximation basically
differs only in the amplitude of the oscillation, being larger for the
latter. 

\indent{As}
a more realistic application, we consider the case where the trap is not
displaced ($z_0 = 0$) and the scattering lengths are equal to the
experimental values as in Table 1 of reference \cite{Wi99}.
In fig. 1 we show the results of the calculation corresponding to the frozen model,
full-model and its linearization, when $\phi (0) = 0.7 \pi$ as
explained in the caption. 
As shown in these figures, there is again complete agreement between
the ``full'' and ``frozen''model. 
The translational degrees of freedom remain at its equilibrium values,
whereas the displacements of the ones associated to the ``breathing''
oscillation are negligible (notice the scale of the figs. 1d and 1e). 
Therefore, we can conclude that the invariant subspace found in the
symmetrical case survives and, as a consequence, we have a complete
agreement between the two models.

\indent{In} 
the case under investigation, the parameters of the frozen model are
$\omega_R = 0.100$ and $\Lambda=-0.017$, showing that we are in the
``strong Josephson'' coupling regime $\frac{2\omega_{R}}{|\Lambda|}>1$
\cite{Timmermans98,Ci98}. 
The linearization of the frozen model gives a frequency equal to
$\omega_F=\sqrt{\omega_{R}(\omega_{R}+\frac{\Delta}{2})}$
which coincides with the lowest energy normal mode of equation (13),
whose frequency  is $\omega_{nm}^{(1)}=0.096$. 
As in the symmetrical case, 
the linear approximation of the full model differs from the other two
models in the amplitude of the oscillation, with the former
having a larger amplitude.
Of course, as $\theta (0)$ approaches the equilibrium value, 
$\theta (0)=\pi$, this difference is less pronounced.

\indent{Now} we will turn to the investigation of the effect of the
displacement of the traps on the system.
To this end, we repeat the previous calculation with $z_0=0.15 z_{sho}$
\cite{Wi99}.
In fig. 2 we show the curves for the relative population fraction,
translation and ``breathing''oscillation of each 
condensate component in the three models, with $\phi (0)=0.7\pi$ as
explained in the captions.

\indent{From}
these figures we see that there is a good agreement between the
full-model and its linear approximation whereas they both differ from
the ``frozen'' model.
From figure (2a) we observe that, for the relative population, the
``frozen'' model gives a much smaller amplitude and a much higher
frequency, compared  to the other two models. 
One the other hand, the results of the calculation according to the full model
and its linear approximation are very close, the latter having a
slighter higher amplitude and frequency. 
We observe the same qualitative pattern for the translational degrees
of freedom, which oscillate in phase and the ``breathing'' ones, which
oscillate out of phase. 

\indent{In} 
the present case, the parameters of the frozen model are
$\Delta=128.4$ and $\omega_R=0.0523$, showing that we are in the
regime of weak Josephson coupling $\frac{2\omega_R}{|\Delta|} < 1$. 
The decrease of $\omega_{R}$ and the increase of $\Delta$ can be
understood as basically an effect of the displacement of the traps.
This displacement diminishes the overlap between the condensate
orbitals $\phi_j(\vec r,{\bf 0})$, which decreases $\omega_R$ and
$\bar\lambda_{12}$ and increases $\Delta$. 

\indent{The} 
linear approximation of the frozen model gives a frequency
of $\omega_F\approx 1.832$ much higher than the observed
oscillation of the relative population.
On the other hand, the lowest energy normal mode of the linear
equations (13) has a frequency equal to $0.4466$, which compares
well with the frequency of oscillation for the relative population and
the translation degree of freedom calculated in the linear
approximation to the full-model.
This result indicates that these degree of freedom has components basically
in this normal mode. In turn, the time-evolution of the
``breathing''variables, figs. (2d) and (2e), although dominated by
this mode, show the presence of higher frequency modes.

\indent{Comparing}
the two calculations at different displacements, we see that the full
model predicts that, when we displace the trap, the amplitude
diminishes and the frequency increases, in agreement with reference
\cite{Wi99}.
The frozen model gives the same qualitative prediction, however it
overestimates the size of the effect. 

\indent{To} summarize, we have investigated the influence of the change
of the quantum state in the tunneling dynamics of Josephson coupled
two-component BEC, for different trap geometries.
From our calculations, we have observed that when the trap is not
displaced the tunneling dynamics is ``coherent'', that is, the
orbitals of each condensate do not change.
As a consequence, we have agreement between the frozen model
\cite{Sm97,RSK99} and the full model. 
On the other hand, when we displace the traps, the change in the
condensate orbitals has a strong influence in the tunneling dynamics.
Compared to the frozen model, which neglects this change, the full
model gives a smaller frequency and higher amplitude of oscillation.
We also found that in this latter case the full-model is much closer to
its linear approximation than to the frozen model.

\indent{As}
a final remark, it is clear that, by changing the initial condition,
we can get a large variety of dynamical behaviors for the system.
Our aim in this paper was to establish a framework where all these
questions could be investigated in a scheme numerically feasible and
which can be easily extended to a three dimensional calculation. 
The application of the formalism presented in this paper to this more
realistic case is currently under investigation.    

\bigskip

ACKNOWLEDGMENT: 
This work was partially supported
by Fundac\c{c}\~ao de Amparo $\grave{a}$ Pesquisa do Estado de
S$\tilde{a}$o Paulo (FAPESP).
E. J. V. de Passos was supported in part by CNPq. 
The work of Da-Shin Lee was supported in part by the National Science
Council, ROC under the Grant NSC-89-2112-M-259-008-Y.

\begin{center}
{\bf Figure Caption}
\end{center}

\baselineskip=18pt

\noindent{Figure} 1. Plot showing time evolution of relative
population (1a), translation (1b-1c) and breathing (1d-1e).  
The solid curve corresponds to the full
model, the doted curve to the ``frozen'' model and dashed to the linear
approximation.
The initial condition is the relative phase at $\theta(0)=0.7\pi$ and the other
variables are at its equilibrium values.
The displacement of the trap $z_0$ is set equal to 0. The time is
expressed in unit of $\frac{1}{\omega_0}$ and the lengths in unit of
$z_{sho}$, which is the size parameter of the trap. See text for more
details. 

\bigskip

\noindent{Figure} 2. The conventions are the same as in figure 1. 
The displacement of the trap $z_0$ is set equal to $0.15z_{sho}$.
The initial phase is $\theta(0)=0.7\pi$ and the other
variables are at its equilibrium values.
See text for more details. 

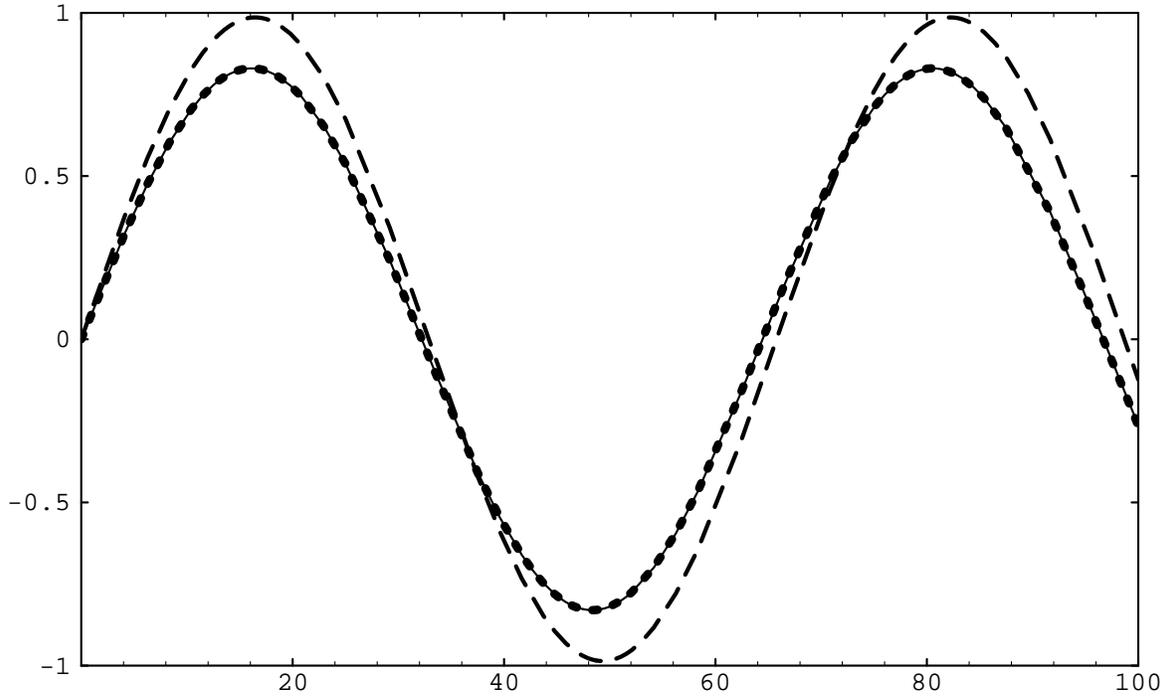
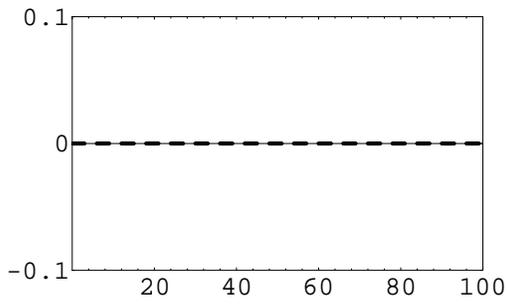
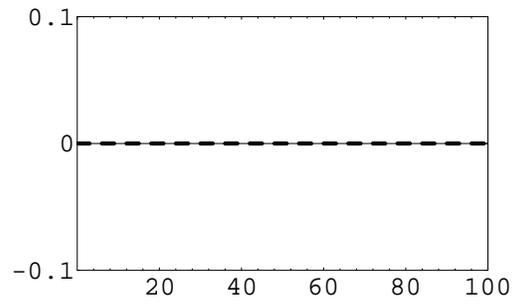
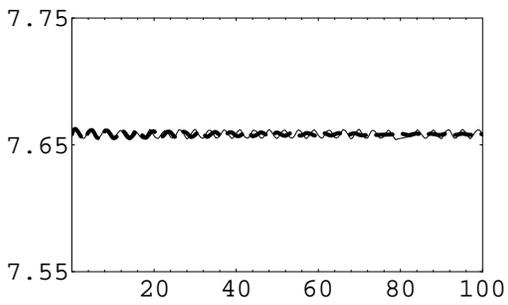
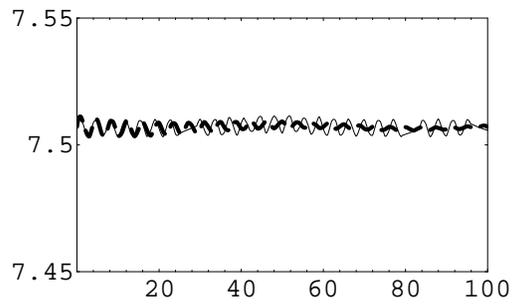
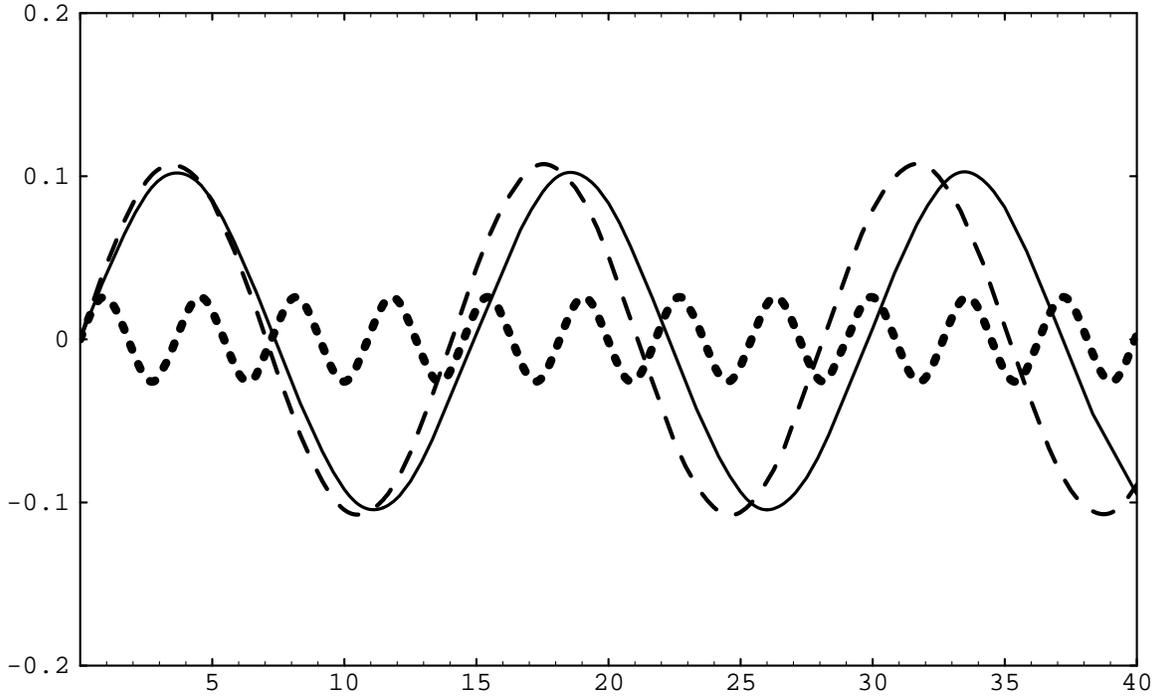
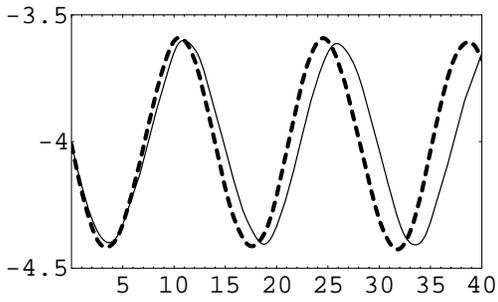
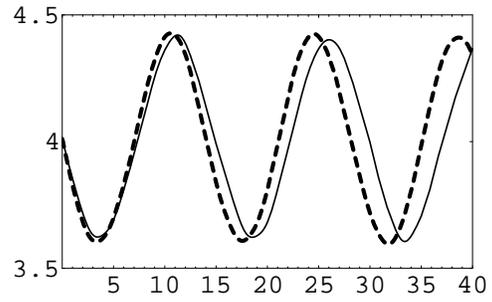
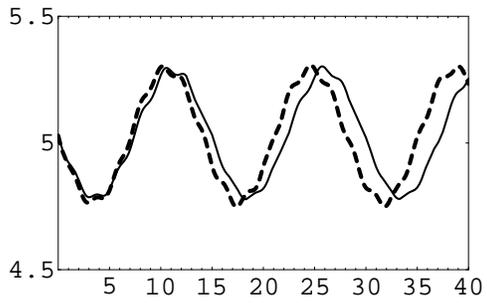
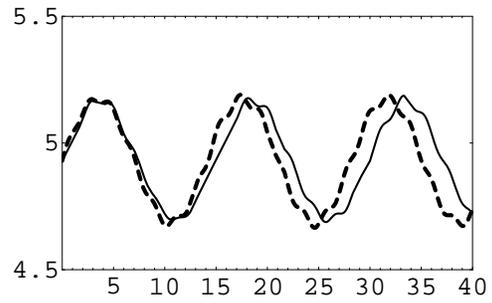
\end{document}